\documentclass[%
reprint,
showpacs,
amsmath,amssymb,
aps,
pra,
]{revtex4-1}

\usepackage[dvipsnames]{xcolor}
\usepackage{graphicx}
\usepackage{dcolumn}
\usepackage{bm}
\usepackage{lipsum}
\usepackage{float}
\usepackage{dsfont}
\usepackage{subfigure}

\usepackage[colorlinks=true,bookmarks=false]{hyperref}
\hypersetup{linkcolor=blue,citecolor=red,urlcolor=blue}   

\begin{document}

\preprint{APS/123-QED}

\title{Second order kinetic Kohn-Sham lattice model}

\author{S.Sol\'{o}rzano} \email{sosergio@ethz.ch} \affiliation{ ETH
  Z\"urich, Computational Physics for Engineering Materials, Institute
  for Building Materials, Wolfgang-Pauli-Str. 27, HIT, CH-8093 Z\"urich
  (Switzerland)}

\author{M. Mendoza} \email{mmendoza@ethz.ch} \affiliation{ ETH
  Z\"urich, Computational Physics for Engineering Materials, Institute
  for Building Materials, Wolfgang-Pauli-Str. 27, HIT, CH-8093 Z\"urich
  (Switzerland)}

\author{H. J. Herrmann}\email{hjherrmann@ethz.ch} \affiliation{ ETH
  Z\"urich, Computational Physics for Engineering Materials, Institute
  for Building Materials, Wolfgang-Pauli-Str. 27, HIT, CH-8093 Z\"urich
  (Switzerland)}



\begin{abstract}
In this work we introduce a new semi-implicit second order correction scheme to the kinetic Kohn-Sham lattice model. The new approach is validated by performing realistic exchange-correlation energy calculations of atoms and dimers of the first two rows of the periodic table finding good agreement with the expected values. Additionally we simulate the ethane molecule where we recover the bond lengths and compare the results with standard methods. Finally, we discuss the current applicability of pseudopotentials within the lattice kinetic Kohn-Sham approach.
\end{abstract}

\pacs{31.15.-p, 47.11.Qr, 31.15.X- }
\maketitle

Central to many theoretical and practical problems in molecular and condensed matter physics, quantum chemistry, and material science is the solution of the Schr\"{o}dinger equation for systems of electrons in molecules and crystals. It is well known that this problem scales exponentially \cite{RevModPhys.71.1253} in the number of electrons and is in general a non trivial task. This has led to the development of a number of approximate solution methods with different degrees of success in different scenarios: density matrix renormalisation group (DMRG) methods\cite{PhysRevLett.69.2863} for 1D model systems, Hartree Fock\cite{PhysRev.35.210.2,fock}, quantum Monte Carlo\cite{RevModPhys.73.33}, exact diagonalization, and Kohn-Sham density functional theory (DFT)\cite{PhysRev.136.B864,PhysRev.140.A1133}.  

The DFT formalism has proven to be one of the most versatile methods for ground state electronic calculations in spite of its known shortcomings\cite{:/content/aip/journal/jcp/136/15/10.1063/1.4704546}. This is due to the fact that it is formally exact, and that it allows to calculate several physical quantities such as bond lengths, bonding energies, ionization energies, etc.  More  specifically, within the Kohn-Sham DFT theory the many-electron Schr\"{o}dinger equation is mapped to an auxiliary problem of non-interacting electrons subject to an external potential $V_{xc}[\rho]$ that  depends on the electron density $\rho$.  $V_{xc}[\rho]$ can be obtained from a universal, albeit, unknown energy functional of the electronic density, and in practice several highly accurate approximations have been proposed over the years.

Recently \cite{PhysRevLett.113.096402} it was shown that the Kohn-Sham equations that describe a system of electrons, can be recovered from an underlaying kinetic model described by the Boltzmann equation. This connection opens the door to the possibility of including exchange and correlation corrections into electronic calculations based on a kinetic perspective. Furthermore, it also allows the use of efficient Lattice Boltzmann (LB) methods\cite{LBW} to solve the Kohn-Sham equations. LB methods, are well known numerical tools in the area of computational fluid dynamics. In recent years, however, their use has been extended beyond fluids to fields such as quantum mechanics\cite{Succi1993327}, relativistic hydrodynamics\cite{PhysRevLett.105.014502} and classical electrodynamics\cite{PhysRevE.82.056708} among others. The success of the LB method in these different fields is due to its high flexibility and computational efficiency that stems from the local character of the LB equations.

Here we present a new semi-implicit second order correction scheme to the lattice kinetic Kohn-Sham approach. The scheme avoids both the computational load of solving systems of equations with implicit methods and the instabilities associated to explicit corrections. This is achieved by using the formally known time evolution of a general state in imaginary time. The scheme is validated by performing calculations of the exchange and correlation energies of small atoms, as well as bond length calculations of $\text{H}_2$, LiH and $\text{C}_{2}\text{H}_{6}$ molecules. We also report on how pseudopotentials couple to our scheme, as well as the effects of using different lattices. 

This work is organized as follows: section \ref{method} revises the kinetic formulation of the Kohn-Sham equations, then in section \ref{LKM} the lattice kinetic model is explained, and the semi-implicit second order correction to the forcing term is presented. Afterwards in section \ref{results} the proposed model is studied and validated using, as a benchmark, calculations of exchange-correlation energies of different atoms and bond lengths of simple dimers. Results regarding the use of pesudopotentials and some comments on the choice of different lattices are also presented. Finally, in section \ref{conclusion} the conclusions and future work are discussed.


\section{Kinetic approach to DFT\label{method}}

The kinetic approach to density functional theory is based on the observation that the time dependent Kohn-Sham equations in imaginary time can be recovered as a special macroscopic limit of the Boltzmann equation in the BGK approximation \cite{PhysRev.94.511}, i.e,
\begin{equation}
\frac{\partial f}{\partial t}+\mathbf{v}\cdot\nabla f =-\frac{1}{\tau_{k}}\left(f-f^{eq}\right)+S,
\label{eq.Boltzmann}
\end{equation}
where $\tau_{k}$ is the kinetic relaxation time, $f(\mathbf{x},\mathbf{v},t)$ is a distribution function in phase space, $f^{eq}(\mathbf{x},\mathbf{v},t)$ is the equilibrium distribution of the considered system and $S(\mathbf{x},\mathbf{v},t)$ is a general source term. Given the equilibrium function $f^{eq}$, the macroscopic limit of Eq.\eqref{eq.Boltzmann} is a hierarchy of conservation-like laws that involves only the moments (macroscopic fields) of $f^{eq}$ and that describes their dynamics. The specification of the equilibrium function depends on the system under consideration, thus $f^{eq}$ can be taylored such that its moments follow a prescribed dynamics by imposing constrains on them. For instance if a Maxwellian is used, classical fluids can be described, but a Juttner distribution is required for relativistic systems. In the present case, the target is diffusive dynamics, let the $\alpha_{1}\alpha_{2}...\alpha_{n}$ component of the $n$-th moment of the distribution function, equilibrium distribution function, and source term be respectively defined as
\begin{subequations}
\begin{align}
\Pi^{(n)}_{\alpha_{1}\alpha_{2}...\alpha_{n}}&=\int \mathbf{dv} v_{\alpha_{1}}v_{\alpha_{2}}...v_{\alpha_{n}}f(\mathbf{x},\mathbf{v},t), \label{eq.MomentsDef1} \\
\Pi^{(n)eq}_{\alpha_{1}\alpha_{2}...\alpha_{n}}&=\int \mathbf{dv} v_{\alpha_{1}}v_{\alpha_{2}}...v_{\alpha_{n}}f^{eq}(\mathbf{x},\mathbf{v},t),\\
\Sigma^{(n)}_{\alpha_{1}\alpha_{2}...\alpha_{n}}&=\int \mathbf{dv} v_{\alpha_{1}}v_{\alpha_{2}}...v_{\alpha_{n}}S(\mathbf{x},\mathbf{v},t),
\end{align}
\label{eq.GenMoments}
\end{subequations}
where $v_{\alpha}$ is the $\alpha$ component of the phase space velocity vector $\mathbf{v}$. If the moments of the equilibrium distribution function and source term are chosen as
\begin{subequations}
\begin{align}
&\Pi^{(0) eq}=\Pi^{(0)}, \label{eq.MomentsDif1}\\
&\Pi^{(1) eq}=0, \label{eq.MomentsDif33} \\
&\Pi^{(2) eq}_{ij}=C_{s}^{2}\Pi^{(0)}\delta_{ij},\label{eq.MomentsDif44}\\
&\Pi^{(n) eq}=0 \text{  for  } n>2,\\
&\Sigma^{0}=\tilde{S},\\
&\Sigma^{(n) eq}=0 \text{  for  } n>0,\label{eq.MomentsDif2}
\end{align}
\label{eq.MomentsDif}\noindent
\end{subequations}
where $C_{s}^{2}$ is a characteristic speed of the system and $\tilde{S}$ is a known function of space and time, then it can be shown that the 0-th moment of the distribution function evolves in time according to the diffusion equation (appendix \ref{Ap1})
\begin{equation}
\frac{\partial\Pi^{0}}{\partial t}=\tau_{k}C_{s}^{2}\nabla^{2}\Pi^{0}+\tilde{S}.
\label{eq.Diffusion}
\end{equation}
If the identifications $\Pi^{0}\equiv\psi$, $\tau_{k}C_{s}^{2}\equiv\frac{\hbar}{2m}$ and $\tilde{S}\equiv-\frac{V}{\hbar}\psi$ are made, then Eq.\eqref{eq.Diffusion} can be rewritten as
\begin{equation}
\frac{\partial\psi}{\partial t}=\frac{\hbar}{2m}\nabla^{2}\psi-\frac{V}{\hbar}\psi,
\label{eq.ImagKS}
\end{equation}
which is the Wick rotated time dependent Kohn-Sham equation provided that $V=V_{ion}+V_{ee}+V_{xc}$ is the total potential felt by the electrons, where $V_{ion}$ is the external Coulomb ionic potential, $V_{ee}$ the electron-electron interaction and $V_{xc}$ the exchange-correlation potential. Notice that Eq.\eqref{eq.ImagKS} contains all the ground state information of the considered system.

In principle, for any initial condition $\psi^{(0)}$ of $\Pi^{(0)}$, that has a non vanishing projection on the ground state $\phi_{0}$, the imaginary time evolution guarantees that as the time increases and the wave function is renormalized, only the $\phi_{0}$ contribution is obtained. 

Assuming that $\phi_{0}$ is known, $\phi_{1}$ can be retrieved by changing the initial condition to $\psi^{(0)}-\langle\phi_{0}|\psi^{(0)}\rangle\phi_{0}$. Higher states can similarly be obtained by sequentially removing the lower ones from the initial condition. This serial form of proceeding can be cast in a parallel version that naturally fits the kinetic approach at the expense of solving various kinetic equations. 

Consider a system for which $N$ Kohn-Sham orbitals $\phi_{l}$ $l=1,2,\dots,N$ are required, each one of them being associated to an extended kinetic model 
\begin{equation}
\frac{\partial f_{l}}{\partial t}+\mathbf{v}\cdot\nabla f_{l} =-\frac{1}{\tau_{k}}\left(f_{l}-f^{eq}_{l}\right)+S+\frac{1}{\tau_{k}}W_{l},
\label{eq.Boltzmannortho}
\end{equation}
where $W_{l}$ is an orthonormalization potential defined by the moments
\begin{subequations}
\begin{align}
&\Omega^{(0)}=-\sum_{i<l}\frac{\langle\psi_{l}|\psi_{i}\rangle} {\langle\psi_{i}|\psi_{i}\rangle} \psi_{i},\\
&\Omega^{(1)}=0,\\
&\Omega^{(2)}=-\tau_{k}C^{2}_{s}\sum_{i<l}\frac{\langle\psi_{l}|\psi_{i}\rangle} {\langle\psi_{i}|\psi_{i}\rangle} \psi_{i},\\
&\Omega^{(n)}=0 \text{  for  } n>2.
\end{align}
\end{subequations}
Since the structure of the moments of $W_{l}$ is the same as that of $f^{eq}_{l}$, each $\Pi^{(0)}_{l}$ will be given by
\begin{equation}
\Pi^{(0)}_{l}=\psi_{l}-\sum_{i<l}\frac{\langle\psi_{l}|\psi_{i}\rangle} {\langle\psi_{i}|\psi_{i}\rangle} \psi_{i}.
\end{equation}
Thus, $\Pi^{(0)}_{k}$ has no contributions from $\psi_{l}$ with $l<k$. Since $\psi_{l}\rightarrow\phi_{l}$ before $\psi_{k}\rightarrow\phi_{k}$ for $l<k$ it follows that effectively $\psi_{k}$ has no components along any of the $\phi_{l}$ eigenstates and thus the next lowest available eigenstate is the one that is going to be selected by the imaginary time evolution. Notice also that as the different orbitals start converging, the effect of $W_{l}$ becomes weaker due to the orthonormality of the wave functions and once the different orbitals have converged it plays no further role. In other terms, $W_{l}$ is only used to drive the different $f_{l}$ in such a way that they converge to different Kohn-Sham orbitals.

\section{Lattice Kinetic Model\label{LKM}}
In order to solve Eq.\eqref{eq.Boltzmann} subject to the constraints Eq.\eqref{eq.MomentsDif} and the implicit requirement that the different orbitals need to be orthonormal, a lattice kinetic model for the Kohn-Sham orbitals was developed \cite{PhysRevLett.113.096402}. First the treatment of Ref.\cite{PhysRevLett.113.096402} is revised and then our improvement in the source term is presented. Without loss of generality the derivations are shown for a single orbital, and the expressions for many orbitals are introduced afterwards. 

\subsection{First order LKKS}

Full details of the passage from Eq.\eqref{eq.Boltzmann} to the Lattice Boltzmann equation Eq.\eqref{eq.LatticeBoltzmann} can be found in Ref.\cite{PhysRevE.56.6811}. However, the main ideas are sketched as follows: Eq.\eqref{eq.Boltzmann} can be formally written as an ordinary diferential equation 
\begin{equation}
\frac{df}{dt}+\frac{f}{\tau_{k}}=\frac{g}{\tau_{k}},
\label{eq.de1}
\end{equation}
where $g=f^{eq}+\tau_{k}S$ and $\frac{d}{dt}=\frac{\partial}{\partial t}+\mathbf{v}\cdot\nabla$ is the time derivative along the characteristic line $\mathbf{v}$. Formal integration of Eq.\eqref{eq.de1} in the time interval $[0,\delta t]$, the assumption that $g$ can be linearly approximated in that interval and neglecting terms of order $O(\delta t^{2})$ lead to 
\begin{equation}
f(\mathbf{x}+\mathbf{v}\delta t,t+\delta t)-f(\mathbf{x},t)=-\frac{\delta t}{\tau_{k}}(f(\mathbf{x},t)-f^{eq}(\mathbf{x},t))+\delta tS.
\end{equation}
The $\mathbf{v}$ space discretization is obtained by requiring that the moments Eq.\eqref{eq.GenMoments} can be exactly evaluated, up to certain order, by quadratures e.g.
\begin{align}
\Pi^{(n)}_{\alpha_{1}\alpha_{2}...\alpha_{n}}&=\int \mathbf{dv} v_{\alpha_{1}}v_{\alpha_{2}}...v_{\alpha_{n}}f(\mathbf{x},\mathbf{v},t) \nonumber \\
&=\sum_{i}^{Q}v_{i,\alpha_{1}}v_{i,\alpha_{2}}...v_{i,\alpha_{n}}W_{i}f(\mathbf{x},\mathbf{v}_{i},t) \nonumber \\
&=\sum_{i}^{Q}v_{i,\alpha_{1}}v_{i,\alpha_{2}}...v_{i,\alpha_{n}}f_{i}(\mathbf{x},t),
\end{align}
where $Q$ is the number of lattice vectors $\mathbf{v}_{i}$ and $v_{i,\alpha_{n}}$ denotes the $\alpha_{n}$ component of the $i$-th velocity vector. Similar expressions hold for the lattice moments of $f^{eq}_{i}$ and $S_{i}$.
The lattice-Boltzmann equation for a distribution function $f(\mathbf{x},\mathbf{v},t)$ is thus given by 
\begin{equation}
f_{i}(\mathbf{x}+\mathbf{v}_{i}\delta t,t+\delta t)-f_{i}(\mathbf{x},t)=-\frac{1}{\tau_{k}}(f_{i}(\mathbf{x},t)-f_{i}^{eq}(\mathbf{x},t))+\delta tS_{i},
\label{eq.LatticeBoltzmann}
\end{equation}
where for simplicity $\frac{\delta t}{\tau_{k}}\rightarrow\frac{1}{\tau_{k}}$ that is, the (numerical) relaxation time in the Lattice Boltzmann equation needs not to be the same as in the Boltzmann equation.



If the lattice moments satisfy the constraints Eq.\eqref{eq.MomentsDif}, it can be shown\cite{PhysRevLett.113.096402} that the 0-th order moment of the lattice distribution function $f_{i}$ evolves according to the diffusion equation 
\begin{equation}
\frac{\partial\Pi^{(0)}}{\partial t}=\delta t\left(\tau_{k}-\frac{1}{2}\right)C^{2}_{s}\nabla^{2}\Pi^{(0)}+\tilde{S}+O(\delta t),
\label{eq.model1}
\end{equation}
which is equivalent to Eq.\eqref{eq.ImagKS}, up to terms of order $O(\delta t)$, if the identifications $\Pi^{0}\equiv\psi$, $\delta t\left(\tau_{k}-\frac{1}{2}\right)C_{s}^{2}\equiv\frac{\hbar}{2m}$ and $\tilde{S}\equiv-\frac{V}{\hbar}\psi$ are made. The extra ``$-1/2$'' term in the identification of $\hbar/2m$ is due to the spacial discretization.
\subsection{Second order LKKS}

 By following the standard procedure to obtain Eq.\eqref{eq.model1} i.e. first Taylor expand the l.h.s of Eq.\eqref{eq.LatticeBoltzmann}, and then perform a Chapman-Enskog multi scale expansion\cite{chapman}, it can be shown that the terms of order $\delta t$ in Eq.\eqref{eq.model1} arise from the source term in Eq.\eqref{eq.LatticeBoltzmann}. The same procedure allows to show that if Eq.\eqref{eq.LatticeBoltzmann} is extended as  
\begin{align}
f_{i}(\mathbf{x}+\mathbf{v}_{i}\delta t,t+\delta t)-f_{i}(\mathbf{x},t)=&-\frac{1}{\tau_{k}}(f_{i}(\mathbf{x},t)-f_{i}^{eq}(\mathbf{x},t)) \nonumber\\
&+\delta tS_{i}+\frac{\delta t^{2}}{2}D_{i}S_{i},
\label{eq.LatticeBoltzmann2}
\end{align}
where $D_{i}=\frac{\partial}{\partial t}+\mathbf{v}_{i}\cdot\nabla$, then the terms of order $\delta t$ can be eliminated. More specifically the Taylor expansion,up to second order of Eq.\eqref{eq.LatticeBoltzmann2} , leads to
\begin{equation}
\delta t D_{i}f_{i}+\frac{\delta t^{2}}{2}D^{2}_{i}f_{i}=-\frac{1}{\tau_{k}}(f_{i}-f^{eq}_{i})+\delta tS_{i}+\frac{\delta t^{2}}{2}D_{i}S_{i},
\label{eq.TexpandedLB}
\end{equation}
Observe that the convective terms of the original Boltzmann equation are already present in the first order terms of Eq.\eqref{eq.TexpandedLB} while the second order ones appear due to the fact that the Lattice-Boltzman equation is a discrete approximation.

In the multi-scale expansion the distribution functions and time derivatives are expanded in a small parameter $\epsilon$ (that in fluids dynamics plays the role of a Knudsen number) as follows
\begin{align}
&f=f^{(0)}+\epsilon f^{(1)}+\epsilon^{2}f^{(2)}+\cdots, \label{eq.CEexpansion1}\\
&\frac{\partial}{\partial t}=\epsilon\frac{\partial}{\partial t_{1}}+\epsilon^{2}\frac{\partial}{\partial t_{2}}+\cdots,
\label{eq.CEexpansion2}
\end{align}
whereas space derivatives and source term are rescaled as $\nabla = \epsilon\nabla_{1}$ and $S=\epsilon S^{(1)}$ respectively. Replacing Eq.\eqref{eq.CEexpansion1} and Eq.\eqref{eq.CEexpansion2} into Eq.\eqref{eq.TexpandedLB}, using Eq.\eqref{eq.MomentsDif1} and collecting terms of equal order in $\epsilon$, we obtain
\begin{align}
&O(\epsilon^{0}):\,\,\,f_{i}^{(0)}=f_{i}^{eq},\label{eq.oe0}\\
&O(\epsilon^{1}):\,\,\,D_{1i}f^{eq}_{i}=-\frac{1}{\tau_{k}\delta t}f^{(1)}_{i}+S^{(1)}_{i},\label{eq.oe1}\\
&O(\epsilon^{2}):\,\,\,\frac{\partial f^{eq}_{i}}{\partial t_{2}}+D_{1i}f^{(1)}_{i}+\frac{\delta t}{2}D^{2}_{1i}f^{eq}_{i}=\label{eq.oe2}\\
&\,\,\,\,\,\,\,\,\,\,\,\,\,\,\,\,\,\,\,\,\,-\frac{1}{\tau_{k}\delta t}f^{(2)}+\frac{\delta t}{2}D_{1i}S_{i}^{(1)}.\nonumber
\end{align}
Eq.\eqref{eq.oe1} is further substituted in the l.h.s of Eq.\eqref{eq.oe2} and the result is summed over all discrete velocities leading to the following relation
\begin{equation}
\frac{\partial \Pi^{0}}{\partial t_{2}}+\left(1-\frac{1}{2\tau_{k}}\right)\nabla_{1}\cdot\sum\mathbf{v}_{i}f^{(1)}_{i}=0. 
\label{eq.CEexpansion3}
\end{equation}

The second term on the l.h.s of Eq.\eqref{eq.CEexpansion3} can be evaluated by taking the product of Eq.\eqref{eq.oe1} with $\mathbf{v}_{i}$ and summing over all velocities, obtaining
\begin{equation}
\frac{\partial \Pi^{(0)}}{\partial t_{2}}=\left(\tau_{k}-\frac{1}{2}\right)C^{2}_{s}\nabla^{2}_{1}\Pi^{(0)}.
\label{eq.CEexpansion4}
\end{equation}
In a similar manner if Eq.\eqref{eq.oe1} is summed over all velocities the result is
\begin{equation}
\frac{\partial \Pi^{(0)}}{\partial t_{1}}=\tilde{S}^{(1)}.
\label{eq.CEexpansion5}
\end{equation}
Eq.\eqref{eq.CEexpansion4} can be multiplied by $\epsilon^{2}$ and added to Eq.\eqref{eq.CEexpansion5} multiplied by $\epsilon$. The result is that Eq.\eqref{eq.LatticeBoltzmann} together with the moments constraints, Eq.\eqref{eq.MomentsDif}, implies that $\Pi^{(0)}$ evolves according to
\begin{equation}
\frac{\partial\Pi^{(0)}}{\partial t}=\left(\tau_{k}-\frac{1}{2}\right)C^{2}_{s}\nabla^{2}\Pi^{(0)}+\tilde{S}+O(\delta t^{2}).
\label{eq.model2}
\end{equation}
The inclusion of the term $\frac{\delta t^{2}}{2}D_{i}S_{i}$ in Eq.\eqref{eq.LatticeBoltzmann2} is not intuitive, however it can be seen that if it had not been added then the additional term $\frac{\delta t}{2}\frac{\partial S^{(1)}}{\partial t_{1}}$ would be present in Eq.\eqref{eq.CEexpansion3} and then propagated to Eq.\eqref{eq.model2}. 
It is important to notice that the inclussion of the correction term is only possible because, as shown before, it does not change the macroscopic limit of the Lattice-Boltzmann equation i.e. the zeroth moment of the distribution function still follows a diffusive dynamics.

The system of coupled equations for many orbitals is simply given by 
\begin{align}
&f_{il}(\mathbf{x}+\mathbf{v}_{i}\delta t,t+\delta t)-f_{il}(\mathbf{x},t)= \nonumber \\
&-\frac{1}{\tau_{k}}(f_{il}(\mathbf{x},t)-f_{ik}^{eq}(\mathbf{x},t))+\frac{\delta t^{2}}{2}D_{i}S_{il}+\delta tS_{il}+\frac{1}{\tau_{k}}W_{il},
\label{eq.LKKS}
\end{align}
where the index $l$ is associated to the $l$-th orbital.

At this point, the form of $f^{eq}$, $S$, and $\mathbf{v}_{i}$ remains undefined. To fix these quantities notice that only the moments of $f^{eq}$ and $S$ are required and not the complete analytical form of the functions. This allows to expand both $f^{eq}$ and $S$ in series of the form $f^{eq}(\mathbf{x},\mathbf{v},t)=w(\mathbf{v})F^{eq}(\mathbf{x},\mathbf{v},t)$ where $w(\mathbf{v})$ is the weight function associated to the family of orthogonal polynomials $\{P_{l}\}$ and $F^{eq}(\mathbf{x},\mathbf{v},t)=\sum_{l}a_{l}(\mathbf{x},t)P_{l}(\mathbf{v})$. The coefficients $a_{l}(\mathbf{x},t)$ are easily calculated from the definition of moments and the fact that any combination $v_{\alpha_{1}}v_{\alpha_{2}}...v_{\alpha_{n}}$ of velocities can be expressed in terms of orthogonal polynomials $P_{l}$. If exact quadrature of the moments $\Pi^{(n)eq}$ is demanded up to a fixed order $n\leq N$ we obtain the system of equations

\begin{equation}
\Pi^{(n)eq}_{\alpha_{1}\alpha_{2}...\alpha_{n}}=\sum_{i}w_{i}v_{i,\alpha_{1}}v_{i,\alpha_{2}}...v_{i,\alpha_{n}}F^{eq}(\mathbf{x},\mathbf{v}_{i},t),
\end{equation}
for the weights $w_{i}$ and velocities $\mathbf{v}_{i}$. The solution of this system is in general non unique, and only solutions for which the velocity vectors form a space-filling lattice and the weights are positive definite should be considered. Once a suitable solution is found the $f^{eq}_{i}$ functions introduced in Eq.\eqref{eq.LatticeBoltzmann} are explicitly given by
\begin{equation}
f^{eq}_{i}=w_{i}\sum_{l=0}^{N-1}a_{l}(\mathbf{x},t)P_{l}(\mathbf{v}_{i}).
\end{equation}
A similar expression holds for $S_{i}$. If Hermite tensor polynomials are used, two known lattices that satisfy constrains up to $N=4$ and $N=6$ are respectively D3Q19 and D3Q111. In these cases $f^{eq}_{i}$ and $S_{i}$  are explicitly given by 
\begin{widetext}
\begin{subequations}
\begin{align}
f_{i}^{(D3Q19)}(\mathbf{x},t)&=\Pi^{0}(\mathbf{x},t)w_{i}\left(1+\frac{1}{2C_{s}^{4}}(C_{s}^{2}-D)(3C_{s}^{2}-\mathbf{v}_{i}\cdot\mathbf{v}_{i})\right),\\
S_{i}^{(D3Q19)}(\mathbf{x},t)&=\tilde{S}(\mathbf{x},t)w_{i}\left(1+\frac{1}{2C_{s}^{4}}(C_{s}^{2})(3C_{s}^{2}-\mathbf{v}_{i}\cdot\mathbf{v}_{i})\right),\\
f_{i}^{(D3Q111)}(\mathbf{x},t)&=\Pi^{0}(\mathbf{x},t)w_{i}\left(1+\frac{1}{2C_{s}^{4}}(C_{s}^{2}-D)(3C_{s}^{2}-\mathbf{v}_{i}\cdot\mathbf{v}_{i})+\frac{1}{8C_{s}^{2}}(C_{s}^{2}-2D)(15C_{s}^{4}-10C_{s}^{2}\mathbf{v}_{i}\cdot\mathbf{v}_{i}+(\mathbf{v}_{i}\cdot\mathbf{v}_{i})^{2})\right),\\
S_{i}^{(D3Q111)}(\mathbf{x},t)&=\tilde{S}w_{i}\left(1+\frac{1}{2C_{s}^{4}}(C_{s}^{2})(3C_{s}^{2}-\mathbf{v}_{i}\cdot\mathbf{v}_{i})+\frac{1}{8C_{s}^{2}}(C_{s}^{2}-2D)(15C_{s}^{4}-10C_{s}^{2}\mathbf{v}_{i}\cdot\mathbf{v}_{i}+(\mathbf{v}_{i}\cdot\mathbf{v}_{i})^{2})\right),
\end{align}
\label{exfunctions}
\end{subequations}
\end{widetext}
where $D=(\tau_{k}-\frac{1}{2})C_{s}^{2}=\frac{\hbar}{2m}$ and the value of $C_{s}$ depends on the lattice.
\subsection{Semi-implicit correction\label{XXX}}
From a computational perspective the implementation of Eq.\eqref{eq.LatticeBoltzmann} (model 1) requires no special discussion as it conforms to standard Lattice Boltzmann schemes. However, in our approach, (model 2) Eq.\eqref{eq.LatticeBoltzmann2},  there are various ways to implement the correction term $D_{i}S_{i}$. These can be explicit
\begin{subequations}
\begin{align}
&D_{i}S_{i}=\frac{1}{\delta t}\left(S_{i}(\mathbf{x},t)-S_{i}(\mathbf{x}-\mathbf{v}_{i}\delta t,t-\delta t)\right),\label{eq.Explicit1a}\\
&D_{i}S_{i}=\frac{1}{\delta t}\left(S_{i}(\mathbf{x}+\mathbf{v}_{i}\delta t,t)-S_{i}(\mathbf{x},t-\delta t)\right),\label{eq.Explicit1b}
\end{align}
\label{eq.Explicit1}
\end{subequations}
or implicit
\begin{equation}
D_{i}S_{i}=\frac{1}{\delta t}\left(S_{i}(\mathbf{x}+\mathbf{v}_{i}\delta t, t+\delta t)-S_{i}(\mathbf{x},t)\right).
\label{eq.Implicit}
\end{equation}
Given that explicit implementations do not require solving a system of equations at every iteration of the algorithm, we performed tests using Eqs.\eqref{eq.Explicit1a} and \eqref{eq.Explicit1b}, in both cases we found that the procedure was numerically unstable leading to wild oscillations of the measured quantities. To use the implicit form of $D_{i}S_{i}$ and avoid the necessity of solving systems of equations, Eq.\eqref{eq.Implicit} was approximated using the formally known imaginary time evolution of the different orbitals. That is, for the $n$-th orbital $D_{i}S_{i,n}=D_{i}\frac{V}{\hbar}\psi_{n}$, and its discretized version is given by
 \begin{equation}
D_{i}S_{i,n}=\frac{1}{\delta t}\left(\frac{V}{\hbar}\psi_{n}(\mathbf{x}+\mathbf{v}_{i}\delta t, t+\delta t)-\frac{V}{\hbar}\psi_{n}(\mathbf{x},t)\right).
\label{eq.Implicit2}
\end{equation}
An approximation of Eq.\eqref{eq.Implicit2} can be obtained if $\psi_{n}(\mathbf{x}, t+\delta t)$ can be estimated. From the imaginary time evolution it is known that 
\begin{equation}
\psi_{n}(\mathbf{x},t)=\sum_{j\geq n}c_{j}\phi_{j}e^{-\frac{\epsilon_{j}t}{\hbar} }, 
\end{equation}
where $\phi_{j}$ and $\epsilon_{j}$ are the eigenfunctions and eigenenergies of the Kohn-Sham Hamiltonian and $c_{j}$ are the projection coefficients of $\psi^{0}$ in the basis $\{\phi_{j}\}$. Therefore, we consider the approximation
\begin{equation}
\psi_{n}(\mathbf{x},t+\delta t)\approx \psi_{n}(\mathbf{x},t)e^{-\frac{\epsilon_{n}\delta t}{\hbar}}.
\label{approx}
\end{equation}
Notice that this approximation improves after every iteration and is exact once the steady state has been reached, this follows from the time projection technique that progressively drives all the $c_{j}\to 0$ for $j\neq n$.

Finally it is worth noticing that the actual implentation of the correction term does not add any extra complexity to the scheme. It correspond to an extra scalar-matrix-vector multiplication of the same kind used to calculate the original source term.

\section{results\label{results}}

\subsection{Model Comparison}

Our improved scheme, using the semi-implicit correction term, was used to calculate the exchange and correlation energies of H, He, Be and Ne atoms as well as the bond lengths of $\text{H}_{2}$ and LiH dimers. We compared it with Model 1 using a D3Q19 lattice and the BLYP exchange correlation potential\cite{PhysRevA.38.3098,PhysRevB.37.785} . The physical length of the simulation box is given by $L_{p}=L\Delta x$ where $L$ is the number of grid points in one direction, $\Delta x$ is the distance between two successive sites and the resolution of the system is defined as $\Delta x^{-1}$. 

\begin{figure}[h!]
\includegraphics[scale=0.65]{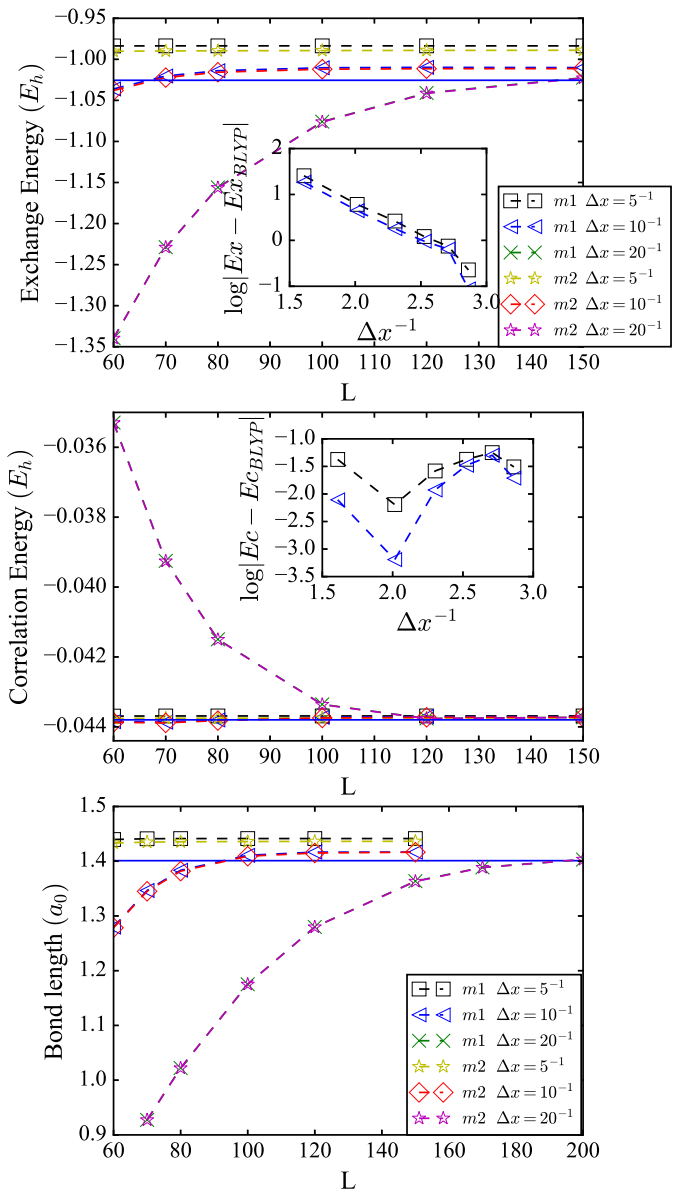}
\caption{Model Comparison: the top and medium panels are respectively the calculated exchange and correlation energies of the He atom as a function of system size for three different resolutions using models 1 and 2. The insets show the relative error as a function of $\Delta x^{-1}$ for a system size of $L=150$. The bottom panel shows the equilibrium length of the $\text{H}_{2}$ molecule. The solid blue line is the DFT result using the BLYP functional reported in Ref.\cite{DFTM,NISTH2BL} }
\label{fig1}
\end{figure}

Results from the He atom and $\text{H}_{2}$ molecule are shown in Fig.\ref{fig1}. Both models show that for a fixed resolution $\Delta x^{-1}$, as the number of lattice sites i.e the physical size of the simulation box increases, the value of the measured quantities tend to converge to a limiting value, and as the resolution level is improved, the limiting value approaches the expected BLYP values Ref.\cite{DFTM,NISTH2BL}. These two behaviors are consistent with the fact that as the physical system size and resolution increase, the boundary effects are reduced and the system better approximates an atom or molecule in free space. Furthermore, the limiting values of model 2 are closer to the expected BLYP values than those of model 1 for a fixed $\Delta x$, and as the resolution improves both models tend to agree. These observations were consistently verified for all the other studied systems.

\begin{figure}
\includegraphics[scale=0.55]{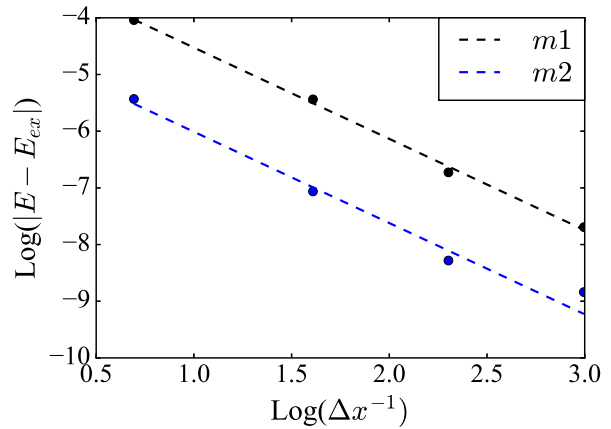}
\caption{Relative error of the ground state energy of the H atom as a function of the resolution for m1 and m2. The circles correspond to the simulation data and the dashed lines are the linear fits. In both cases $\Delta E_{H}\propto\Delta x^{1.6}$}
\label{figR1}
\end{figure}

The degree to which model 2 is more accurate than model 1 with respect to the expected BLYP values depends on the considered atom or molecule and measured quantity ($E_{x}$, $E_{c}$, bond length, etc). For example, the insets in the first and second panels of Fig.\ref{fig1} show respectively the relative error, $\Delta E_{x}$ and $\Delta E_{c}$,  of the exchange and correlation energies as a function of $\Delta x^{-1}$ when the system size is fixed at $L=150$. In the case of $E_{x}$, $\Delta E_{x}\propto\Delta x^{1.5}$ for the first model and $\Delta E_{x}\propto\Delta x^{1.6}$ for the second. In contrast the behavior of $\Delta E_{c}$ is non monotonic. It is worth noting that $\Delta E_{x}$ ranges from $4\%$ to less than $1\%$ whereas $\Delta E_{c}$ is always smaller than $0.3\%$.At this point the difference between the two models seems small, however this is due to the fact that $E_{x}$ and $E_{c}$ are the integrals of non trivial functions of the density and the density gradient, where the later has to be numerically calculated. To better observe the difference between both models, the ground state energy of the H atom, which only requires the norm of the wave function at two consecutive time steps, was calculated. Its relative error $\Delta E_{H}$ as a function of the resolution is shown in Fig.\ref{figR1}. Where it can be observed that $\Delta E_{H}$ in the m2 model is one order of magnitude smaller than in the m1 model. That is an indication that m2 indeed solves the kinetic Kohn-Sham equation more accurately than m1. Finally the results for simulations of the other systems are summarized in Table.\,\ref{tab:table1}.

\begin{table}
\begin{ruledtabular}
\begin{tabular}{lcccc}
Atom& $-E_{x}$ & $-E_{x}$ BLYP & $-E_{c}$ & $-E_{c}$ BLYP\\
\hline
H  & -0.301   & -0.301   & &  \\
He & -1.0197  & -1.0255  & -0.0437 & -0.0438\\
Be & -2.6741  & -2.6578  & -0.0965 & -0.0945\\
Ne & -12.0532 & -12.1378 & -0.3827 & -0.3835\\
\hline
Atom& Bond length & Bond length (BLYP)& & \\
\hline
H2 & 1.3867 & 1.4000 & \\
LiH & 3.005 & 3.016 & \\
\end{tabular}
\end{ruledtabular}
\caption{\label{tab:table1} Exchange and correlation energies of different atoms, and bond lengths of different molecules calculated using model 2 with D3Q19 lattice compared to the known BLYP values.}
\end{table}

\subsection{Ethane molecule}
\begin{figure}
\includegraphics[scale=0.5]{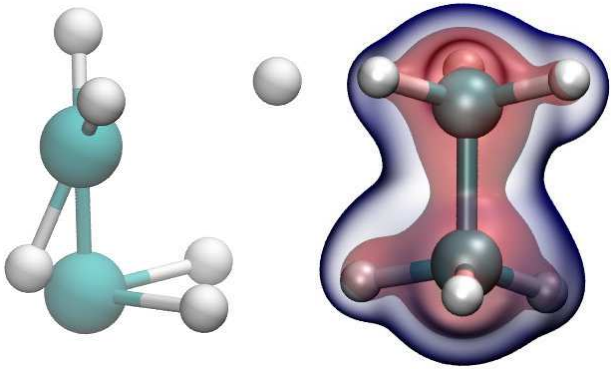}
\caption{(Color online). On the left side the initial configuration of the atoms that conform the ethane molecule is depicted. On the right side the final configuration is shown along with the electron density, the regions of high(low) electron density are indicated by red(blue) surfaces}
\label{figfancy2}
\end{figure}

As a test of the proposed model, the $\text{C}_{2}\text{H}_{6}$ (ethane) molecule was simulated. The carbon atoms were initially located such that their center of mass was in the center of the simulation box and they were aligned along the $z$ axis. The H atoms were randomly located, three of them closer to the upper carbon atom, and the remaining ones closer to lower carbon atom (Fig\,\ref{figfancy2} left). This set up mimics the common scenario in which there is only partial information available. The final configuration, obtained after 3.6 days of run time on a single core, is shown in Fig\,\ref{figfancy2}where the qualitatively correct shape of the ethane molecule and electronic density distribution can be observed, compared to the initial configuration. The relative errors of the bond lengths and angles with respect to the expected ones\cite{NISTH2BL} are $1.3\%$ for the $C-C$ bond length, a mean relative error of $2.1\%$ for the $H-C$ bond length and a $7\%$ for the $H-C-H$ angles. Except for the angles, the accuracy is comparable to that of a Carr-Parinello Molecular dynamics (CPMD) simulation performed with identical initial conditions using a wavefunction cutoff of 100$Ry$. It achieves a $2.7\%$ $C-C$ bond length error, $2.0\%$ mean $H-C$ bond length error and a $0.05\%$ mean $H-C-H$ angle error. The CPMD simulation took about three hours, which is a small fraction of the computational time spent by our model. However CPMD uses pseudopotentials while our model considers the bare Coulomb potential. 

\subsection{Pseudopotentials}

Pseudopotentials are a way to reduce the computational cost of atomistic simulations that works under the approximation that core electrons are mostly inert\cite{CPHC:CPHC201100387} and play a minimal role in most of the chemistry. Although pseudopotentials are designed to be highly accurate and transferable, it is not always clear, a priori, how they do couple to different simulation methods. For instance pseudopotentials are known to be problematic or not directly applicable in diffusion Monte Carlo and Green functions approaches \cite{:/content/aip/journal/jcp/132/15/10.1063/1.3380831,:/content/aip/journal/jcp/95/5/10.1063/1.460849,PhysRevLett.61.1631}.

In order to asset how pseudopotentials couple to our method, tests were perfomed using the dual-space Gaussian pseudopotentials (DSGPP) introduced in Ref.\cite{PhysRevB.54.1703}. The DSGPP were chosen, because they are optimized for the BLYP exchange correlation potential used in this work and because their real space representation is compatible with the real space nature of our method. 

\begin{figure}
\includegraphics[scale=0.61]{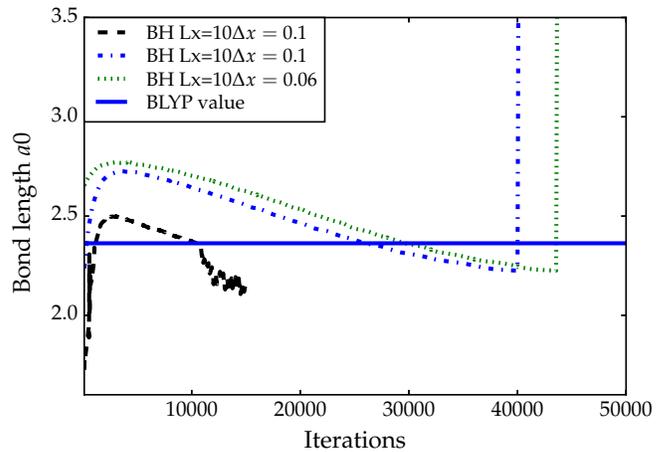}
\caption{(Color online). BH bond length calculated using pseudopotentials. The different lines show that the use of pseudopotentials within our approach leads to numerical instabilities.}
\label{figBH}
\end{figure}

The first non trivial example that includes both local and non local contributions of the DSGPP is the BH molecule where the two inner electrons of the boron atom are neglected. In this case the B atom is described by the pseudopotential $V_{pp}=V_{\text{loc}}+H_{\text{nonloc}}$ where 
\begin{align}
&V_{\text{loc}}(\mathbf{r})=\frac{-Z_{ion}e}{|\mathbf{r}-\mathbf{R}_{B}|}\text{erf}\left(\frac{|\mathbf{r}-\mathbf{R}_{B}|}{\sqrt{2}r_{loc}}\right)+\text{exp}\left(\frac{|\mathbf{r}-\mathbf{R}_{B}|^{2}}{2r_{loc}^{2}}\right)\\ \nonumber
&\times\left[C_{1}+C_{2}\frac{|\mathbf{r}-\mathbf{R}_{B}|^{2}}{r_{loc}^{2}}+C_{3}\frac{|\mathbf{r}-\mathbf{R}_{B}|^{4}}{r_{loc}^{4}}+C_{4}\frac{|\mathbf{r}-\mathbf{R}_{B}|^{6}}{r_{loc}^{6}}\right],
\end{align}
and
\begin{align}
H_{\text{nonloc}}(\mathbf{r},\mathbf{r}')&=\sum_{i=1}^{2}Y_{0,0}(\hat{\mathbf{r}})p_{i}^{0}(\mathbf{r})h_{i}^{0}p_{i}^{0}(\mathbf{r}')Y_{0,0}^{*}(\hat{\mathbf{r}}')\\ \nonumber
&+\sum_{m}Y_{1,m}(\hat{\mathbf{r}})p_{1}^{1}(\mathbf{r})h_{1}^{1}p_{1}^{1}(\mathbf{r}')Y_{1,m}^{*}(\hat{\mathbf{r}}'). 
\end{align}
The values of the constants $C_{i}\,\,i=1,..4$, $r_{loc}$ and $h_{i}$ as well as the functional form of the projectors $p_{i}^{0}(\mathbf{r})$,$p_{1}^{1}(\mathbf{r})$ can be found in Ref.\cite{PhysRevB.54.1703}. The DSGPP for boron was implemented and used to calculate the bond length of the BH molecule. It was found that when pseudopotentials are used within our approach the scheme becomes unstable (black dashed line Fig.\ref{figBH}). The instabilities were partially controlled by artificially resetting the wave functions and electron density to their initial values after a fixed number of iterations while keeping the current position of the ions (blue dashed-dot line Fig.\ref{figBH}), but eventually instabilities arise. Different initial conditions and resolution levels also suffer from instabilities (green doted line Fig.\ref{figBH}). The reason why our approach becomes unstable may be related to the overall nonlinear nature of the system and the nonlocal component of the pseudopotential that requires the evaluation of projection integrals of the form $\int p_{i}^{0}(\mathbf{r}')Y_{l,m}^{*}(\hat{\mathbf{r}}')\psi(\mathbf{r}')$, that may not be sufficiently resolved due to the fact that $\psi(\mathbf{r}')$ is only known at a limited number of lattice points.

\subsection{Lattice Performance}

\begin{figure}
\includegraphics[scale=0.61]{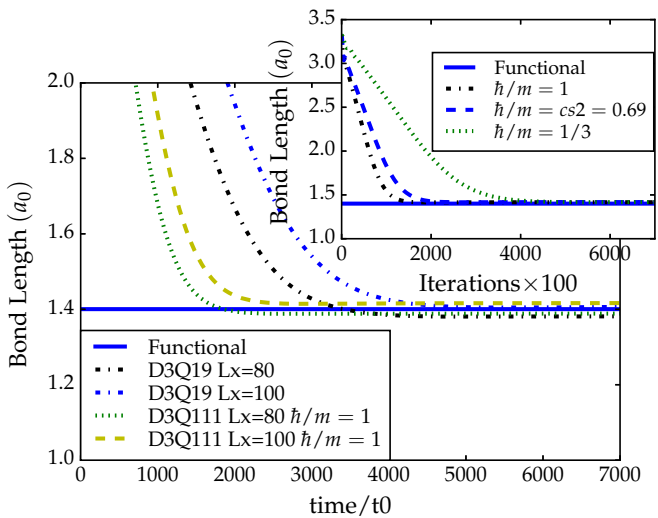}
\caption{Effect of the numerical value of the diffusivity $\hbar/m$ on the convergence speed using the D3Q111 lattice in the simulation of the  $\text{H}_{2}$ molecule.}
\label{fig5}
\end{figure}
The performance and accuracy of model 2 using either the D3Q19 or D3Q111 lattice was investigated by simulating the He atom as well as $\text{H}_{2}$ and LiH dimers. The generally observed trend is that for small resolutions the results obtained using both lattices differ, but as the resolution increases the difference is reduced and the results converge. 

Although both lattices lead to practically the same results, the D3Q111 lattice allows to chose a higher value for the diffusivity $\hbar/m$, that can be used to control the convergence rate of the procedure. As an example, the $\text{H}_{2}$ molecule was simulated using the D3Q111 lattice for three different values of $\hbar/m$ Fig.\ref{fig5}(inset). It can be seen that in all cases the system converges to the same value of the bond length, but for $\hbar/m=1$ the approach is faster than for $\hbar/m=1/3$ or $\hbar/m=0.69$. The D3Q111 lattice has almost six times more velocity vectors than D3Q19, and the time of a single iteration using the D3Q111 lattice was measured to be about 1.5 times longer than that of the D3Q19 lattice. After accounting for this, the comparison between the speed of convergence using both lattices for two different system sizes is presented in Fig.\ref{fig5}. It can be observed that the use of the D3Q111 lattice allows for a faster convergence. However, for large resolutions, since both lattices lead to the same accuracy, using D3Q111 presents no advantage in terms of computational time. For instance for the Be atom we found that the D3Q111 lattice with  $\hbar/m=1$ converges equally fast as D3Q19. 
\section{Conclusions\label{conclusion}}
In this work a new and more accurate Lattice Boltzmann scheme to solve the kinetic Kohn-Sham equations has been introduced and validated. The scheme uses a novel way of implementing a semi implicit second order corrections to the forcing term, that makes use of the known asymptotic behavior of the simulated orbitals. This approach avoids the instabilities of the explicit implementations and the computational load of solving implicit systems of equations. 

The use of pseudopotentials within our approach requires further work to eliminate the associated instabilities and computational demands, not only for the tested case, but also for general pseudopotentials. Possible approaches in that direction include subgrid refinements. 

The results of the ethane molecule simulation show that our method can reproduce the bond lengths of complex molecules, but that further work is required to achieve an overall performance similar to that of established methods such as CPMD, including the full integration with pseudopotentials. It was also confirmed that the D3Q19 and D3Q111 lattices lead to the same results for high enough resolution, giving an advantage to the D3Q19 lattice in terms of computational resources. 

For future work, in addition to the aforementioned improvements for pseudopotentials, the performance of our approach will be investigated in crystal systems where its real space periodic boundary conditions naturally fit.

\begin{acknowledgments}
We acknowledge financial support form the European Research Council (ERC) Advanced Grant 319968-FlowCCS. The authors also thank Sauro Succi for useful discussions.
\end{acknowledgments}

\bibliographystyle{apsrev4-1}
\bibliography{bibliography.bib}

\clearpage
\appendix
\section{Macroscopic difussion limit of the Boltzman equation}\label{Ap1}
In order to show that Eq.\eqref{eq.MomentsDif} leads to a difussive macroscopic behavior we use the Chapman-Enskog procedure\cite{chapman}.
Let the the distribution function and time derivative be expanded as
\begin{align}
&f=f^{(0)}+\epsilon f^{(1)}+\epsilon^{2}f^{(2)}+\cdots, \label{Aem1}\\
&\frac{\partial}{\partial t}=\epsilon\frac{\partial}{\partial t_{1}}+\epsilon^{2}\frac{\partial}{\partial t_{2}}+\cdots,
\end{align}
and let the spatial derivative and source term be rescaled as $\nabla = \epsilon\nabla_{1}$ and $S=\epsilon S^{(1)}$, where $\epsilon$ is regarded as a small quantity. Substituting these relations in Eq.\eqref{eq.Boltzmann} and equating terms of equal order in $\epsilon$ the following set of equations are found
\begin{align}
&f^{(0)}=f^{eq}, \label{Ae0}\\
&\partial_{t_{1}}f^{(0)}+\mathbf{v}\cdot\nabla_{1}f^{(0)}=-\frac{1}{\tau_{k}}f^{(1)}+S^{(1)}, \label{Ae1} \\
&\partial_{t_{1}}f^{(1)} + \partial_{t_{2}}f^{(0)}+\mathbf{v}\cdot\nabla_{1}f^{(1)}=-\frac{1}{\tau_{k}}f^{(2)}. \label{Ae2}
\end{align}
Taking the 0-th moment of Eq.\eqref{Ae1} and \eqref{Ae2} we found
\begin{align}
&\partial_{t_{1}}\Pi^{(0,0)}+\nabla_{1}\cdot\Pi^{(1,0)}=-\frac{1}{\tau_{k}}\Pi^{(0,1)}+\Sigma^{(0,1)},\\
&\partial_{t_{1}}\Pi^{(0,1)}+\partial_{t_{2}}\Pi^{(0,0)}+\nabla_{1}\cdot\Pi^{(1,1)}=\frac{1}{\tau_{k}}\Pi^{(0,2)}.
\end{align}
Where $\Pi^{(i,j)}$ is the i-th moment of $f^{(j)}$. and $\Sigma^{(0,1)}$ is the 0-th moment of the rescaled source term. Eq. \eqref{Aem1}, \eqref{Ae0} and the constrain Eq.\ref{eq.MomentsDif1} implies that the previous equations simplify as 
\begin{subequations}
\begin{align}
&\partial_{t_{1}}\Pi^{(0)}+\nabla_{1}\cdot\Pi^{(1,0)}=\Sigma^{(0,1)},\\
&\partial_{t_{2}}\Pi^{(0)}+\nabla_{1}\cdot\Pi^{(1,1)}=0.
\end{align}
\label{Ae3}
\end{subequations}
Eq.\eqref{Ae0} and constrain Eq.\eqref{eq.MomentsDif33} implies that $\Pi^{(1,0)}=0$. Thus Eq.\eqref{Ae3} can be written as 
\begin{subequations}
\begin{align}
&\partial_{t_{1}}\Pi^{(0)}=\Sigma^{(0,1)},\label{Ae4a}\\
&\partial_{t_{2}}\Pi^{(0)}+\nabla_{1}\cdot\Pi^{(1,1)}=0\label{Ae4b}.
\end{align}
\label{Ae4}
\end{subequations}

$\Pi^{(1,1)}$ can be calculated by taking the first moment of Eq.\eqref{Ae1}, due to the constrains Eq.\eqref{eq.MomentsDif33} and \eqref{eq.MomentsDif2} only the term $\mathbf{v}\cdot\nabla_{1}f^{(0)}$ will contribute, explicetely 
\begin{align}
\Pi^{(1,1)}_{k}&=-\tau_{k}\int v_k \mathbf{v}\cdot\nabla_{1}f^{(0)} d^{3}v \nonumber \\
&=-\tau_{k}\partial_{i}\int v_{k}v_{i}f^{(0)}d^{3}v \nonumber \\
&=-\tau_{k}\partial_{i}\Pi^{(0)}_{i,j}.\label{Ae5}
\end{align}
$\nabla_{1}\cdot\Pi^{(1,1)}$ is then calculated as
\begin{align}
\nabla_{1}\cdot\Pi^{(1,1)}&=\partial_{i}\Pi^{(1,1)}_{i}\\
&=-\tau_{k}\partial_{i}\partial_{k}\Pi^{(0)}_{i,k}\\
&=-\tau_{k}C_{s}^{2}\nabla_{1}\cdot\nabla_{1}\Pi^{(0)}\\
&=-\tau_{k}C_{s}^{2}\nabla^{2}_{1}\Pi^{(0)},
\end{align}
where the constrain Eq.\eqref{eq.MomentsDif44} was used. Eq.\eqref{Ae4b} then reads
\begin{equation}
\partial_{t_{2}}\Pi^{(0)}-\tau_{k}C_{s}^{2}\nabla^{2}_{1}\Pi^{(0)}=0.
\label{Ae6}
\end{equation}
Finally Eq.\eqref{eq.Diffusion} is obtained by multiplying Eq.\eqref{Ae4a} by $\epsilon$ and Eq.\eqref{Ae6} by $\epsilon^{2}$ and adding them taking into acount the scaling of the spatial derivative and source term, as well as the expansion of the temporal derivative.

\end{document}